\documentclass[webpdf,contemporary,large,namedate]{oup-authoring-template}%

\usepackage{algorithmicx}
\usepackage{lmodern}
\usepackage{setspace}

\begin{document}
\journaltitle{TBD}
\DOI{TBD}
\copyrightyear{2024}
\pubyear{2024}
\access{Advance Access Publication Date: Day Month Year}
\appnotes{Preprint}
\firstpage{1}

\title[BWT construction and query]{BWT construction and search at the terabase scale}
\author[1,2,3,$\ast$]{Heng Li\ORCID{0000-0003-4874-2874}}
\address[1]{Department of Data Science, Dana-Farber Cancer Institute, 450 Brookline Ave, Boston, MA 02215, USA}
\address[2]{Department of Biomedical Informatics, Harvard Medical School, 10 Shattuck St, Boston, MA 02215, USA}
\address[3]{Broad Insitute of MIT and Harvard, 415 Main St, Cambridge, MA 02142, USA}
\corresp[$\ast$]{Corresponding author. \href{mailto:hli@ds.dfci.harvard.edu}{hli@ds.dfci.harvard.edu}}

%\received{Date}{0}{Year}
%\revised{Date}{0}{Year}
%\accepted{Date}{0}{Year}

\abstract{
\sffamily\footnotesize
\textbf{Motivation:}
Burrows-Wheeler Transform (BWT) is a common component in full-text indices.
Initially developed for data compression, it is particularly powerful for encoding redundant sequences such as pangenome data.
However, BWT construction is resource intensive and hard to be parallelized,
and many methods for querying large full-text indices only report exact matches or their simple extensions.
These limitations have hampered the biological applications of full-text indices.
\vspace{0.5em}\\
\textbf{Results:}
We developed ropebwt3 for efficient BWT construction and query.
Ropebwt3 indexed 320 assembled human genomes in 65 hours and indexed 7.3 terabases of commonly studied bacterial assemblies in 26 days.
This was achieved using up to 170 gigabytes of memory at the peak without working disk space.
Ropebwt3 can find maximal exact matches and inexact alignments under affine-gap penalties,
and can retrieve similar local haplotypes matching a query sequence.
It demonstrates the feasibility of full-text indexing at the terabase scale.
\vspace{0.5em}\\
\textbf{Availability and implementation:}
\url{https://github.com/lh3/ropebwt3}
}

\maketitle

\section{Introduction}

Although millions of genomes have been sequenced,
the majority of them were sequenced from a small number of species such as human, \emph{E. coli} and \emph{M. tuberculosis}.
As a result, existing genome sequences are highly redundant.
This is how \citet{Hunt2024.03.08.584059} compressed 7.86 terabases (Tb) of bacterial assemblies, also known as AllTheBacteria, into 78.5 gigabytes (GB)
after grouping phylogenetically related genomes~\citep{Brinda:2024aa}.
The resultant compressed files losslessly keep all the sequences but are not directly searchable.
Indexing is necessary to enable fast sequence search.

K-mer data structures are a popular choice for sequence indexing~\citep{Marchet:2021aa}.
They can be classified into three categories.
The first category does not associate k-mers with their positions in the database sequences.
These data structures support membership query or pseudoalignment~\citep{Bray:2016aa},
but cannot reconstruct input sequences or report base alignment.
Sequence search at petabase scale use all such methods~\citep{Edgar:2022aa,Karasikov2020.10.01.322164,Shiryev:2024aa}.
The second category associates a subset of k-mers with their positions.
Upon finding k-mer matches, algorithms in this category go back to the database sequences and perform base alignment.
Most aligners work this way.
However, because the database sequences are not compressed well,
these algorithms may require large space to store them.
The last category keeps all k-mers and their positions.
Algorithms in this category can reconstruct all the database sequences without explicitly storing them.
Nonetheless, although positions of k-mers can be compressed efficiently~\citep{Karasikov:2020aa},
they still take large space.
The largest lossless k-mer index consists of a few terabases~\citep{Karasikov2020.10.01.322164}.

Compressed full-text indices, such as FM-index~\citep{DBLP:conf/focs/FerraginaM00} r-index~\citep{DBLP:conf/soda/GagieNP18,DBLP:journals/tcs/BannaiGI20,DBLP:journals/jacm/GagieNP20}
and move index~\citep{DBLP:conf/icalp/NishimotoT21},
provide an alternative way for fast sequence search~\citep{DBLP:journals/csur/Navarro21}.
The core component of these data structures is often Burrows-Wheeler Transform (BWT; \citealt*{Burrows:1994aa})
which is a lossless transformation of strings.
The BWT of a highly redundant string tends to group symbols in the original string into long runs
and can thus be well compressed.
When we supplement BWT with a data structure to efficiently compute the rank of a symbol in BWT,
we can in theory count the occurrences of a query string in time linear to its length.
FM-index further adds a sampled suffix array to locate substrings,
while r-index uses an alternative method that is more efficient for redundant strings.
Both of them support compression and sequence search at the same time.

BWT-based indices have been used for read alignment~\citep{Langmead:2009aa,Li:2009uq,Li:2009aa},
\emph{de novo} sequence assembly~\citep{Simpson:2012aa},
metagenome profiling~\citep{Kim:2016aa} and
data compression~\citep{Cox:2012ly}.
They have also emerged as competent data structures for pangenome data.
Existing pangenome-focused tools~\citep{Rossi:2022aa,Ahmed:2021aa,Zakeri:2024aa}
use prefix-free parsing for BWT construction~\citep{Boucher:2019aa}.
They require more memory than input sequences and are impractical for huge datasets.
Although ropebwt2 developed by us can construct BWT in memory proportional to its compressed size and is fast for short strings~\citep{Li:2014ab},
it is inefficient for chromosome-long sequences.
OnlineRlbwt~\citep{DBLP:journals/tcs/BannaiGI20} has a similar limitation.
grlBWT~\citep{DBLP:journals/iandc/DiazDominguezN23} is likely the best algorithm for constructing the BWT of similar genomes.
It reduces the peak memory at the cost of large working disk space.
In its current form, the algorithm does not support update to BWT.
We would need to reconstruct the BWT from scratch when new genomes arrive.
BWT construction for highly redundant sequences remains an active research area.

With BWT-based data structures, it is trivial to test the presence of string $P$ in the index,
but finding substring matches within $P$ needs more thought.
Learning from bidirectional BWT~\citep{DBLP:conf/bibm/LamLTWWY09},
we found a BWT constructed from both strands of DNA sequences supports the extension of exact matches in both directions~\citep{Li:2012fk}.
This gave us an algorithm to compute maximal exact matches (MEMs), which was later improved by \citet{DBLP:conf/dlt/Gagie24} for long MEMs.
\citet{DBLP:journals/tcs/BannaiGI20} proposed a distinct algorithm to compute MEMs without requiring both strands.
Matching statistics~\citep{DBLP:journals/algorithmica/ChangL94} and pseudo-matching length (PML; \citealt*{Ahmed:2021aa}) have also been considered.
All these algorithms find exact local matches only.

It is also possible to identify inexact matches.
\citet{DBLP:journals/tcs/KucherovST16} invented search schemes to find semi-global alignment.
With the BWT-SW algorithm, \citet{Lam:2008aa} simulated a suffix trie, a tree data structure, with BWT
and performed sequence-to-trie alignment to find all local matches
between a query string and the BWT of a large genome.
We went a step further by representing the query string with its direct acyclic word graph (DAWG; \citealt*{DBLP:journals/eatcs/BlumerBEHM83})
and performed DAWG-to-trie alignment.
This is the BWA-SW algorithm~\citep{Li:2010fk}.
Nonetheless, we later realized the formulation of BWA-SW had theoretical flaws
and its implementation was closer to DAWG-to-DAWG alignment than to DAWG-to-tree alignment.

In this article, we will explain our solution to BWT construction
and to sequence search at the terabase scale.
Our contribution includes:
a) a reinterpreted BWA-SW algorithm that fixes issues in our earlier work~\citep{Li:2010fk};
%b) the description of core data structures in ropebwt2~\citep{Li:2014ab} and fermi~\citep{Li:2012fk} that were published before but without technical details;
b) its application in finding similar haplotypes;
c) an incremental in-memory BWT construction implementation that scales to large pangenome datasets;
d) a faster algorithm to find alignment positions with a standard FM-index of highly redundant strings.

\section{Methods}

In a nutshell, ropebwt3 computes the partial multi-string BWT of a subset of sequences with libsais (\url{https://github.com/IlyaGrebnov/libsais})
and merges the partial BWT into the existing BWT run-length encoded as a B+-tree~\citep{Li:2014ab}.
It repeats this procedure until all input sequences are processed.
The BWT by default includes input sequences on both strands.
This enables forward-backward search~\citep{Li:2012fk} required by accelerated long MEM finding~\citep{DBLP:conf/dlt/Gagie24}.
Ropebwt3 also reports local alignment with affine-gap penalty using a revised BWA-SW algorithm~\citep{Li:2010fk}.

Ropebwt3 combines and adapts existing algorithms and data structures.
Nonetheless, the notations here differ from our early work (e.g. from 1-based to 0-based coordinates) and from other publications.
We will describe our methods in full for completeness.

\begin{table}[!tb]
\caption{Notations and naming convention\label{tab:sym}}
\begin{tabular*}{\columnwidth}{@{\extracolsep\fill}ll@{\extracolsep\fill}}
\toprule
Notation & Description \\
\midrule
$\Sigma$   & Alphabet of symbols. $\{{\tt A},{\tt C},{\tt G},{\tt T},{\tt N}\}$ for DNA \\
$\Sigma'$  & Augmented alphabet: $\Sigma'\triangleq\Sigma\cup\{\$\}$ \\
$a,b,c$    & Symbols in $\Sigma'$ \\
$T$        & Concatenated reference string including sentinels \\
$S(i)$     & Suffix array: offset of the $i$-th smallest suffix \\
$B$        & BWT string: $B[i]\triangleq T[S(i)-1]$ \\
$m$        & Number of sentinels: $m\triangleq|\{i:B[i]=\$\}|$ \\
$n$        & Total length: $n\triangleq|T|=|B|$ \\
$r$        & Number of runs: $r\triangleq|\{i:B[i]\not=B[i+1]\}|+1$ \\
$C_B(a)$   & Accumulative count: $C_B(a)\triangleq|\{i:B[i]<a\}|$ \\
${\rm rank}_B(a,k)$ & Rank: ${\rm rank}_B(a,k)\triangleq|\{i<k:B[i]=a\}|$ \\
$\pi_B(a,k)$& $\pi_B(a,k)\triangleq C_B(a)+{\rm rank}_B(a,k)$\\
$\pi(k)$   & LF-mapping: $\pi(k)\triangleq S^{-1}(S(k)-1)=\pi_B(B[k],k)$ \\
\botrule
\end{tabular*}
\end{table}

\subsection{Basic concepts}

\begin{figure}[bt]
\centering
\includegraphics[width=.49\textwidth]{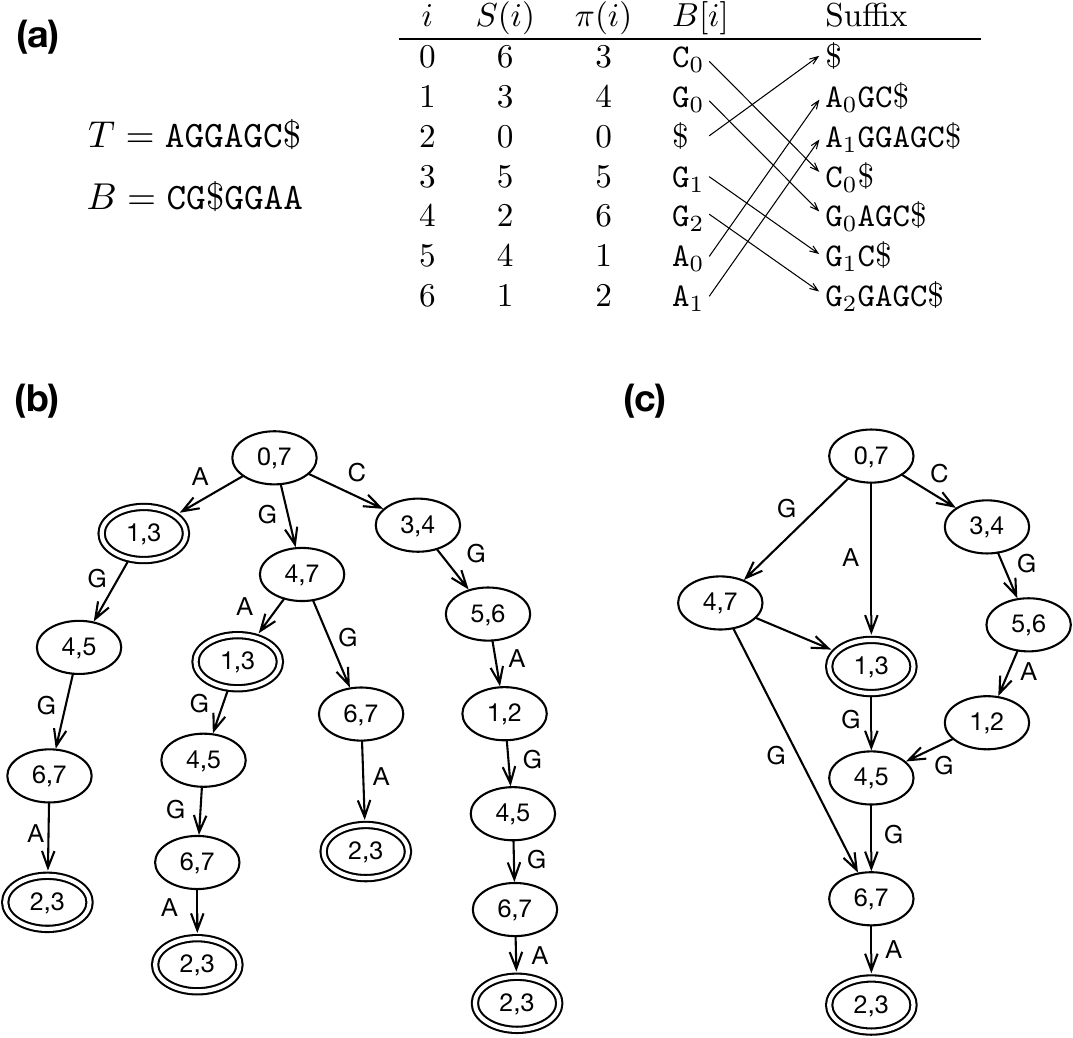}
\caption{Examples of BWT and related data structures.
{\bf (a)} The BWT $B$, suffix array $S$ and LF-mapping $\pi$ of string $T$.
Subscriptions are equal to the ranks of symbols in $B$, which are the same as the ranks among the suffixes (indicated by arrows).
{\bf (b)} The prefix trie simulated with BWT $B$.
In each node, the pair of integers gives the suffix array interval of the string represented by the path from the node to the root.
Double circles indicate nodes that can reach the beginning of $T$.
{\bf (c)} The prefix directed acyclic word graph (DAWG) of $T$ by merging nodes with identical suffix array intervals.}\label{fig:1}
\end{figure}

Let $\Sigma$ be an alphabet of \emph{symbols} (Table~\ref{tab:sym}).
Given a string $P$ over $\Sigma$, $|P|$ is its length and $P[i]\in\Sigma$, $0\le i<|P|$, is the $i$-th symbol in $P$,
and $P[i,j)$ is a substring of length $j-i$ starting at $i$.
Operator ``$\circ$'' concatenates two strings or between strings and symbols.
It may be omitted if concatenation is apparent from the context.

Suppose $\mathcal{T}=(P_0,P_1,\ldots,P_{m-1})$ is an ordered list of $m$ strings over $\Sigma$.
$T\triangleq P_0\$_0P_1\$_1\cdots P_{m-1}\$_{m-1}$ is the concatenation of the strings in $\mathcal{T}$
with sentinels ordered by $\$_0<\$_1<\cdots<\$_{m-1}$.
For other ways to define string concatenation on ordered string lists or unordered string sets, see \citet{Cenzato:2024ab}.

For convenience, let $n\triangleq|T|$ and $T[-1]=T[n-1]$.
The \emph{suffix array} of $T$ is an integer array $S$ such that $S(i)$,
$0\le i<n$, is the start position of the $i$-th smallest suffix among all suffixes of $T$ (Fig.~\ref{fig:1}a).
The \emph{Burrows-Wheeler Transform} (\emph{BWT}) of $T$ is a string $B$ computed by $B[i]=T[S(i)-1]$.
All sentinels in $B$ are represented by the same symbol ``$\$$'' and are not distinguished from each other.
$\Sigma'\triangleq\Sigma\cup\{\$\}$ denotes the alphabet including the sentinel.

For $a\in\Sigma'$, let $C_B(a)\triangleq|\{i:B[i]<a\}|$ be the number of symbols smaller than $a$
and ${\rm rank}_B(a,k)\triangleq|\{i<k:B[i]=a\}|$ be the number of $a$ before offset $k$ in $B$.
We may omit subscription $B$ when we are describing one string only.
The last-to-first mapping (\emph{LF mapping}) $\pi$ is defined by $\pi(i)\triangleq S^{-1}(S(i)-1)$,
where $S^{-1}$ is the inverse function of suffix array $S$.
It can be calculated as $\pi(i)=\pi_B(B[i],i)$, where $\pi_B(a,i)\triangleq C(B[i])+{\rm rank}(B[i],i)$.
As $B[\pi(i)]$ immediately proceeds $B[i]$ on $T$, we can use $\pi$ to decode the $i$-th sequence in $B$.

%\begin{algorithm}[tb]
%	\caption{A simplified bwa-aln algorithm with PSSM}\label{algo:bwa-aln}
%	\begin{algorithmic}[1]
%		\Procedure{BwaAlnPSSM}{$B,P,s,g$}\Comment{$g$ is gap penalty}
%			\State $H\gets\mbox{empty heap prioritized on the 1st value}$
%			\State $H.{\rm push}(0,|P|,0,|T|)$
%			\While{$H$ is not empty}
%				\State $(p,i,l,h)\gets H.{\rm pop}()$\Comment{element with the smallest $p$}
%				\If{$i=0$}
%					\State \Return $[l,h)$\Comment{this is the best match}
%				\EndIf
%				\State $H.{\rm push}(p+g_{i-1},i-1,l,h)$\Comment{insertion}
%				\State $a\gets P[i-1]$
%				\For{$c\in\Sigma$}\Comment{match, mismatch or deletion}
%					\State $l'\gets C_B(c)+{\rm rank}_B(c,l)$\Comment{backward search}
%					\State $h'\gets C_B(c)+{\rm rank}_B(c,h)$
%					\If{$l'<h'$}\Comment{$[l',h')$ is a child of $[l,h)$}
%						\State $H.{\rm push}(p+s_i(a|c),i-1,l',h')$\Comment{(mis)match}
%						\State $H.{\rm push}(p+g_i,i,l',h')$\Comment{deletion}
%					\EndIf
%				\EndFor
%			\EndWhile
%			\State \Return $\emptyset$\Comment{no match}
%		\EndProcedure
%	\end{algorithmic}
%\end{algorithm}

%\begin{algorithm}[!b]
%	\caption{Retrieve the $i$-th sequence, $0\le i<m$}\label{algo:get}
%	\begin{algorithmic}[1]
%		\Procedure{Retrieve}{$B,i$}
%			\State $P\gets\epsilon$\Comment{empty string}
%			\While{$B[i]\not=\$$}
%			\State $P\gets B[i]\circ P$
%			\State $i\gets C_B(B[i])+{\rm rank}(B[i],i)$\Comment{i.e. $i\gets\pi(i)$}
%			\EndWhile
%			\State \Return{$P$}
%		\EndProcedure
%	\end{algorithmic}
%\end{algorithm}

\subsection{BWT construction}

\begin{algorithm}[bt]
	\caption{Insert BWT $B_2$ into BWT $B_1$}\label{algo:merge}
	\begin{algorithmic}[1]
		\Procedure{AppendBWT}{$B_1,B_2$}
			\State $m_1\gets|\{k:B_1[k]=\$\}|$
			\State $m_2\gets|\{k:B_2[k]=\$\}|$
			\For{$k\gets 0$ {\bf to} $|B_2|-1$}
				\State $a\gets B_2[k]$
				\State $R(k)\gets (a,\pi_{B_2}(a,k))$
			\EndFor
			\For{$i\gets 0$ {\bf to} $m_2-1$}
				\State $k\gets i$; $l\gets m_1$
				\Repeat
					\State $(a,k')\gets R(k)$
					\State $R(k)\gets(a,k+l)$\Comment{position in the merged BWT}
					\State $k\gets k'$; $l\gets \pi_{B_1}(a,l)$
				\Until{$a=\$$}
			\EndFor
			\For{$(a,k)\in R$}\Comment{N.B. $k$ is sorted in array $R$}
				\State ${\rm insert}_{B_1}(a,k)$\Comment{insert $a$ after $k$ symbols in $B_1$}
			\EndFor
		\EndProcedure
	\end{algorithmic}
\end{algorithm}

Libsais is an efficient library for computing the suffix array of a single string.
It does not directly support a list of strings.
Nonetheless, we note that $T$ is a string over alphabet $\Sigma''=\{\$_0,\$_1,\ldots,\$_{m-1},{\tt A},{\tt C},{\tt G},{\tt T},{\tt N}\}$
with lexicographical order $\$_0<\cdots<\$_{m-1}<{\tt A}<{\tt C}<{\tt G}<{\tt T}<{\tt N}$.
We can use $m+5$ non-negative integers to encode $T$ and apply libsais.
The suffix array derived this way will be identical to the suffix array of $T$.

\begin{figure}[bt]
\centering
\includegraphics[width=.49\textwidth]{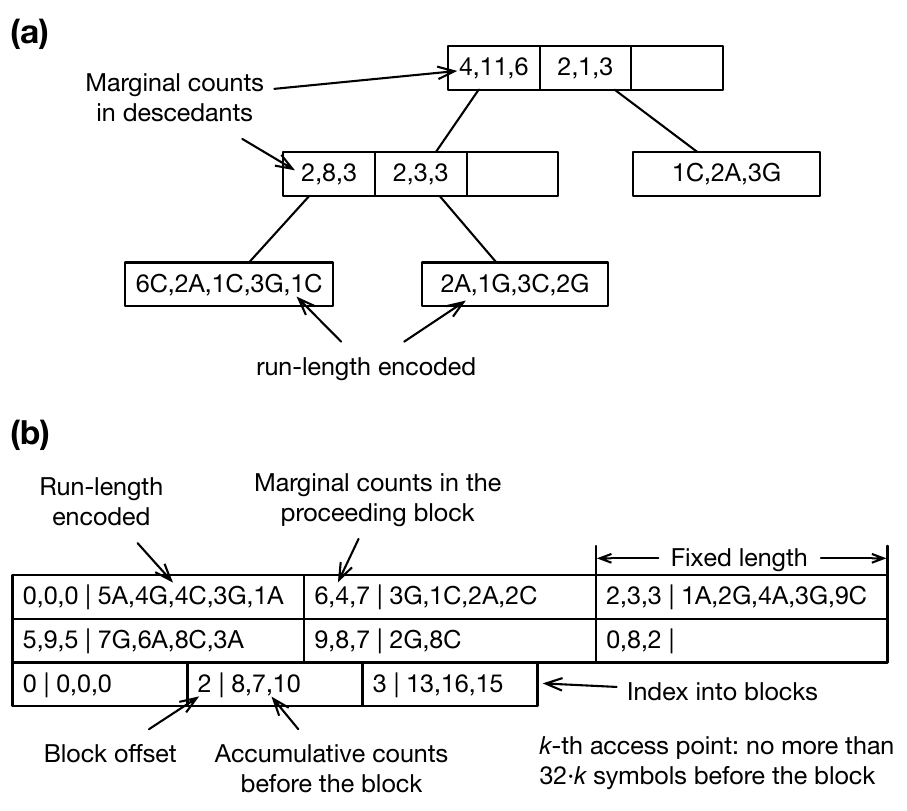}
\caption{Examples of binary BWT encoding.
The alphabet is $\{{\tt A},{\tt C},{\tt G}\}$.
{\bf (a)} Encoded as a B+-tree.
An internal node keeps the marginal counts of symbols in its descendants.
An external node keeps the run-length encoded substring in BWT.
A run length may be encoded in one, two, four or eight bytes in a scheme inspired by UTF-8.
A B+-tree organized this way resembles the rope data structure.
{\bf (b)} Encoded as a bit-packed array.
The first two rows store run-length encoded BWT interleaved with marginal counts in each block.
The Elias delta encoding is used to represent run lengths.
The last row is an index into BWT for fast access.
}\label{fig:2}
\end{figure}

For $m$ that can be represented by a 32-bit integer and $n$ represented by a 64-bit integer,
libsais will need at least $12n$ bytes to construct the suffix array.
It is impractical for $n$ more than tens of billions.
To construct BWT for large $n$, ropebwt3 uses libsais to build the BWT for a batch (up to 7 Gb by default)
and merges it to the BWT of already processed batches (Algorithm~\ref{algo:merge}).
The basic idea behind the algorithm is well known~\citep{DBLP:conf/latin/FerraginaGM10}
but implementations vary~\citep{DBLP:conf/dcc/Siren16,DBLP:conf/dcc/Oliva0SMKGB21}.
In ropebwt3, we encode BWT with a B+-tree (Fig.~\ref{fig:2}a).
This yields $O(\log r)$ rank query (line 13) and insertion (line 17), where $r$ is the number of runs in the merged BWT.
The bottleneck of the algorithm lies in rank calculation (line 8), which can be parallelized if $m_2>1$.

This online BWT construction algorithm does not use temporary disk space.
The overall time complexity is $O(n\log r)$.
The BWT takes $O(r\log r)$ in space.
The memory required for partial BWT construction with libsais depends on the batch size and the longest string.
B+-tree is dynamic.
Ropebwt3 optionally converts B+-tree to the fermi binary format (Fig.~\ref{fig:2}b; \citealt*{Li:2012fk})
which is static but is faster to query and can be memory-mapped.

Ropebwt2~\citep{Li:2014ab} uses the same B+-tree to encode BWT and has the same time complexity.
However, it inserts sequences, not partial BWTs, into existing BWT.
It cannot be efficiently parallelized for long strings.
Note that independent of our earlier work, \citet{DBLP:journals/jda/OhnoSTIS18} also used a B+-tree to encode BWT.
Its implementation~\citep{DBLP:journals/tcs/BannaiGI20} is several times slower than ropebwt2 on 152 bacterial genomes from \citet{Li:2024ab},
possibly because ropebwt2 is optimized for the small DNA alphabet.

\subsection{Suffix array interval and backward search}

For a string $P\in\Sigma^*$ (i.e. not including sentinels), let ${\rm occ}(P)$ be the number of occurrences of $P$ in $T$.
Define ${\rm lo}(P)$ to be the number of suffixes that are lexicographically smaller than $P$
and ${\rm hi}(P)\triangleq {\rm lo}(P)+{\rm occ}(P)$.
$[{\rm lo}(P),{\rm hi}(P))$ is called the \emph{suffix array interval} of $P$, or \emph{SA interval} in brief.
An SA interval is \emph{important} if there exists $P$ such that it is the SA interval of $P$.
The SA interval of the empty string is $[0,n)$ where $n=|T|$.

If we know the SA interval of $P$, we can calculate the SA interval of $aP$ with:
${\rm lo}(aP)=\pi_B(a,{\rm lo}(P))$ and ${\rm hi}(aP)=\pi_B(a,{\rm hi}(P))$.
%\begin{eqnarray*}
%{\rm lo}(aP)&=&C(a)+{\rm rank}(a,{\rm lo}(P))\\
%{\rm hi}(aP)&=&C(a)+{\rm rank}(a,{\rm hi}(P))
%\end{eqnarray*}
To calculate ${\rm occ}(P)$,
we start with $[0,n)$ and repeatedly apply the equation above from the last symbol in $P$ to the first.
This procedure is called \emph{backward search}.

\subsection{Locating with FM-index}

By definition of BWT, if $i\in[{\rm lo}(P),{\rm hi}(P))$, $P=T[S(i),S(i)+|P|)$.
We can thus locate an occurrence of $P$ in the original string $T$.
However, suffix array $S$ takes $O(n\log n)$ in space and may be too large to store explicitly.
With an FM-index, we only store $S(i)$ if and only if $i$ is a multiple of $s$ where $s$ is a positive integer controlling the sample rate.
We calculate the rest of $S(i)$ using LF-mapping $\pi(\cdot)$.

A complication with multi-string BWT is that $\pi(i)$ does not point to the preceeding symbol when $B[i]=\$$.
This is because sentinels are ordered during suffix sorting but are not distinguished from each other in BWT.
We will have to store the rank of each string in $T$ (array $R$ in Algorithm~\ref{algo:locate}).
For convenience, we also keep the index of the string and the offset on the string instead of the offset on the concatenated string $T$.

To find the position of $i\in[{\rm lo}(P),{\rm hi}(P))$ in the original string,
we repeatedly apply $\pi(\cdot)$ $k$ times until the position of $\pi^k(i)$ is stored.
With $U$, $V$ and $R$ precalculated by {\sc IndexSSA} in Algorithm~\ref{algo:locate},
we can locate $i$ with
$$
\left\{\begin{array}{ll}
(R(\pi^k(i)),k) & (B[\pi^k(i)]=\$) \\
(U(\pi^k(i)),V(\pi^k(i)+k)) & (\mbox{otherwise and $\pi^k(i)\bmod s=0$})\\
\end{array}\right.
$$
where $k$ is the smallest non-negative integer such that $B[\pi^k(i)]=\$$ or $\pi^k(i)\bmod s=0$.
On average, $k\approx s$, which means each locate operation triggers $s$ rank queries.

Function {\sc Locate1} in Algorithm~\ref{algo:locate} provides a faster way to find one position in SA interval $[l,h)$.
A key observation is that if the interval contains a sentinel or there exists $k$ such that $l\le ks<h$,
we can immediately locate one occurrence; if not, we can apply backward search repeatedly until an SA interval $[l',h')$
brackets a stored suffix array value.
If $T$ consists of $m'$ identical strings, we apply $s/\min(h-l,m')$ rounds of backward searches on average,
usually much faster than the naive algorithm.
Ropebwt3 implements a generalized {\sc Locate1} function that finds multiple occurrences.

\begin{algorithm}[tb]
	\caption{Locate one hit given suffix array samples}\label{algo:locate}
	\begin{algorithmic}[1]
		\Procedure{IndexSSA}{$B,s$}
			\State $A\gets\emptyset$; $m\gets|\{i:B[i]=\$\}|$\Comment{Number of sequences}
			\For{$t\gets 0$ {\bf to} $m-1$}\Comment{Traverse all sequences}
				\State $k\gets t$; $l\gets 0$\Comment{$l$ will be the length of $t$-th sequence}
				\Repeat\Comment{Iterate from the end of the sequence}
					\State $k\gets\pi(k)$; $l\gets l+1$
					\If{$B[k]=\$$}
						\State $R(k)\gets t$\Comment{The rank of $t$-th sequence is $k$}
					\ElsIf{$k \bmod s=0$}
						\State $A\gets A\cup\{k/s\}$
					\EndIf
				\Until{$B[k]=\$$}
				\For{$k\in A$}
					\State $U(k)=t$; $V(k)\gets l-1-k$
				\EndFor
			\EndFor
			\State \Return $(U,V,R)$
		\EndProcedure
%		\Procedure{Locate}{$B,U,V,R,{\rm lo},{\rm hi},s,l$}
%			\State $m\gets|\{i:B[i]=\$\}|$; $A\gets\emptyset$
%			\If{${\rm lo}={\rm hi}$}
%				\State\Return{$\emptyset$}
%			\ElsIf{${\rm lo}<{\rm hi}\le m$}
%				\For{$k\gets{\rm lo}$ {\bf to} ${\rm hi}-1$}
%					\State $A\gets A\cup\{(R(k),l)\}$
%				\EndFor
%			\ElsIf{$\exists k$ such that $k\cdot s+m\in[{\rm lo},{\rm hi})$}
%				\State $A\gets A\cup\{(U(k),V(k)+l)\}$
%				\State $A\gets A\cup\mbox{\sc Locate}(B,U,V,R,{\rm lo},ks+m,l+1)$
%				\State $A\gets A\cup\mbox{\sc Locate}(B,U,V,R,ks+m+1,{\rm hi},l+1)$
%			\Else
%				\For{$a\in\Sigma\cup\{\$\}$}
%					\State ${\rm lo}'\gets C(a)+{\rm rank}_B(a,{\rm lo})$
%					\State ${\rm hi}'\gets C(a)+{\rm rank}_B(a,{\rm hi})$
%					\State $A\gets A\cup\mbox{\sc Locate}(B,U,V,R,{\rm lo}',{\rm hi}',l+1)$
%				\EndFor
%			\EndIf
%			\State\Return {$A$}
%		\EndProcedure
		\Procedure{Locate1}{$B,U,V,R,{\rm lo},{\rm hi},s$}
			\State $I\gets\{({\rm lo},{\rm hi},0)\}$
			\While{$I\not=\emptyset$}
				\State $(l,h,o)\gets\mbox{largest interval in $I$}$
				\State $I\gets I\setminus \{(l,h,o)\}$
				\If{$\exists k$ such that $k\cdot s\in[l,h)$}
					\State \Return $(U(k),V(k)+o)$
				\ElsIf{$\pi_B(\$,l)<\pi_B(\$,h)$}
					\State \Return $(R(\pi_B(\$,l)),o)$
				\EndIf
				\For{$a\in\Sigma$}
					\State $I\gets I\cup\{(\pi_B(a,l),\pi_B(a,h),o+1)\}$
				\EndFor
			\EndWhile
		\EndProcedure
	\end{algorithmic}
\end{algorithm}

\subsection{Double-strand BWT}\label{sec:ds-bwt}

The definitions above are applicable to generic strings.
With one BWT, we can only achieve backward search;
forward search additionally requires the BWT of the reverse strings~\citep{DBLP:conf/bibm/LamLTWWY09}.
Nonetheless, due to the strand symmetry of DNA strings,
it is possible to achieve both forward and backward search with one BWT provided that the BWT contains both strands of DNA strings~\citep{Li:2012fk}.

Formally, a DNA alphabet is $\Sigma=\{{\tt A},{\tt C},{\tt G},{\tt T},{\tt N}\}$.
$\overline{a}$ denotes the Watson-Crick complement of symbol $a\in\Sigma$.
The complement of $\$$, ${\tt A}$, ${\tt C}$, ${\tt G}$, ${\tt T}$ and ${\tt N}$
are $\$$, ${\tt T}$, ${\tt G}$, ${\tt C}$, ${\tt A}$ and ${\tt N}$, respectively.

For string $P$, $\overline{P}$ is its reverse complement string.
The double-strand concatenation of a DNA string list $\mathcal{T}=(P_0,P_1,\ldots,P_{m-1})$ is
$\tilde{T}=P_0\$_0\overline{P}_0\$_1P_1\$_2\overline{P}_1\$_3\cdots P_{m-1}\$_{2m-2}\overline{P}_{m-1}\$_{2m-1}$.
The \emph{double-strand BWT} (\emph{DS-BWT}) of $\mathcal{T}$ is the BWT of $\tilde{T}$.
We note that if $P$ is a substring of $\tilde{T}$,
$\overline{P}$ must be a substring and ${\rm occ}(P)={\rm occ}(\overline{P})$.
The \emph{suffix array bidirectional interval} (\emph{SA bi-interval}) of $P$ is a 3-tuple defined as $({\rm lo}(P),{\rm lo}(\overline{P}),{\rm occ}(P))$.

\begin{algorithm}[tb]
	\caption{Backward and forward extensions with DS-BWT}\label{algo:backfor}
	\begin{algorithmic}[1]
		\Procedure{BackwardExt}{$B,(k,k',s),a$}
			\ForAll{$b<\overline{a}$}\Comment{$b$ can be $\$$}
				\State $k'\gets k'+\big[\pi_B(\overline{b},k+s)-\pi_B(\overline{b},k)\big]$
			\EndFor
			\State $s\gets \pi_B(a,k+s)-\pi_B(a,k)$
			\State $k\gets \pi_B(a,k)$
			\State \Return{$(k,k',s)$}
		\EndProcedure
		\Procedure{ForwardExt}{$B,(k,k',s),a$}
			\State $(k',k,s)\gets${\sc BackwardExt}$(B,(k',k,s),\overline{a})$
			\State \Return{$(k,k',s)$}
		\EndProcedure
	\end{algorithmic}
\end{algorithm}

An SA bi-interval can be extended in both backward and forward directions.
To calculate the SA bi-interval of $aP$, we can use the standard backward search to compute ${\rm lo}(aP)$ and ${\rm occ}(aP)$ from ${\rm lo}(P)$ and ${\rm occ}(P)$.
As to ${\rm lo}(\overline{aP})$, we note that $[{\rm lo}(\overline{aP}),{\rm hi}(\overline{aP}))\subset[{\rm lo}(\overline{P}),{\rm hi}(\overline{P}))$
because $\overline{aP}=\overline{P}\circ\overline{a}$ is prefixed with $\overline{P}$.
We can thus calculate ${\rm lo}(\overline{aP})={\rm lo}(\overline{P})+\sum_{b<\overline{a}}{\rm occ}(\overline{P}b)={\rm lo}(\overline{P})+\sum_{b<\overline{a}}{\rm occ}(\overline{b}P)$.

It is easy to see that if $(k,k',s)$ is the SA bi-interval of $P$, $(k',k,s)$ is the SA bi-interval of $\overline{P}$, and vice versa.
Therefore, we can use the backward extension of $\overline{P}$ to achieve the forward extension of $P$.
Algorithm~\ref{algo:backfor} gives the details.
It simplifies the original formulation~\citep{Li:2012fk}.

\subsection{Finding supermaximal exact matches}

An \emph{exact match} between strings $T$ and $P$ is a 3-tuple $(t,p,l)$ such that $T[t,t+l)=P[p,p+l)$.
A \emph{maximal exact match} (MEM) is an exact match that cannot be extended in either direction.
A MEM may be contained in another MEM on the pattern string $P$.
For example, suppose $T={\tt GACCTCCG}$ and $P={\tt ACCT}$.
MEM $(5,1,2)$ is contained in MEM $(1,0,4)$ on the pattern string.
A \emph{supermaximal exact match} (SMEM) is a MEM that is not contained in other MEMs on the pattern string.
In the example above, only $(1,0,4)$ is a SMEM.
There are usually much fewer SMEMs than MEMs.
\citet{DBLP:conf/dlt/Gagie24} recently proposed a new algorithm to find long SMEMs (Algorithm~\ref{algo:smem})
that is faster than our earlier algorithm~\citep{Li:2012fk}
especially when there are many short SMEMs to skip.
Both algorithms can also find SMEMs occurring at least $s_{\min}$ times~\citep{DBLP:conf/cpm/TatarnikovFKG23}.

\begin{algorithm}[b]
	\caption{Finding SMEMs no shorter than $\ell$~\citep{DBLP:conf/dlt/Gagie24}}\label{algo:smem}
	\begin{algorithmic}[1]
		\Procedure{FindSMEM1}{$\ell,s_{\min},B,P,i$}
			\State {\bf if} $i+\ell>|P|$ {\bf then} \Return{$|P|$}\Comment{Reaching the end of $|P|$}
			\State $(k,k',s)\gets(0,0,|B|)$\Comment{SA bi-interval of empty string}
			\For{$j\gets i+\ell-1$ {\bf to} $i$}\Comment{backward}
				\State $(k,k',s)\gets${\sc BackwardExt}$(B,(k,k',s),P[j])$
				\State {\bf if} $s<s_{\min}$ {\bf then} \Return{$j+1$}
			\EndFor
			\For{$j\gets i+\ell$ {\bf to} $|P|-1$}\Comment{forward}
				\State $(k,k',s)\gets${\sc ForwardExt}$(B,(k,k',s),P[j])$
				\State {\bf if} $s<s_{\min}$ {\bf then} {\bf break}
			\EndFor
			\State $e\gets j$ and output $[i,e)$\Comment{SMEM found}
			\State $(k,k',s)\gets(0,0,|B|)$
			\For{$j\gets e$ {\bf to} $i+1$}\Comment{backward again}
				\State $(k,k',s)\gets${\sc BackwardExt}$(B,(k,k',s),P[j])$
				\State {\bf if} $s<s_{\min}$ {\bf then} \Return{$j+1$}
			\EndFor
		\EndProcedure
		\Procedure{FindSMEM}{$\ell,s_{\min},B,P$}
			\State $i\gets0$
			\Repeat
				\State $i\gets${\sc FindSMEM1}$(\ell,s_{\min},B,P,i)$
			\Until{$i\ge|P|$}
		\EndProcedure
	\end{algorithmic}
\end{algorithm}

\subsection{Finding inexact matches with BWA-SW}

The prefix trie of $T$ is a tree that encodes all the prefixes of $T$ (Fig.~\ref{fig:1}b).
Each edge in the tree is labeled with a symbol in $\Sigma$.
A path from a node to the root spells a substring of $T$.
We can label a node with the SA interval of the string from the node to the root.
When we know the label of a node, we can find the label of its child using backward search.
We can thus simulate the top-down traversal of the prefix trie~\citep{Lam:2008aa}.

As is shown in Fig.~\ref{fig:1}b, different nodes in the prefix trie may have the same label.
If we merge nodes with the same label (Fig.~\ref{fig:1}c), we will get a prefix DAWG~\citep{DBLP:journals/eatcs/BlumerBEHM83}.
For $|T|>1$, the DAWG has at most $2|T|-1$ nodes~\citep{DBLP:conf/icalp/BlumerBEHM84}.
Each node uniquely corresponds to an important SA interval of $T$.

Let $G_T$ denote the prefix DAWG of $T$ and $V(G_T)$ be the set of vertices in $G_T$.
Given a reference string $T$ and a pattern string $P$, we can align $G_T$ and $G_P$ under affine-gap penalty with
$$
\left\{\begin{array}{lll}
E_{uv}&=&\max_{u'\in{\rm pre}(u)}\{\max\{H_{u'v}-q,E_{u'v}\}\}-e\\
F_{uv}&=&\max_{v'\in{\rm pre}(v)}\{\max\{H_{uv'}-q,F_{uv'}\}\}-e\\
G_{uv}&=&\max_{u'\in{\rm pre}(u),v'\in{\rm pre}(v)}\{H_{u'v'}+s(u',u;v',v)\}\\
H_{uv}&=&\max\{G_{uv},E_{uv},F_{uv}\}\\
\end{array}\right.
$$
where $u\in V(G_P)$, $v\in V(G_T)$,
${\rm pre}(u)$ and ${\rm pre}(v)$ are the sets of predecessors in $G_P$ and $G_T$ respectively,
$q$ is the gap open penalty, $e$ is the gap extension penalty,
and $s(u',u;v',v)$ is the match/mismatch score between the symbol labeled on edge $(u',u)$ in $G_P$
and the symbol on $(v',v)$ in $G_T$.
This equation is similar to but not the same as our earlier result~\citep{Li:2010fk}.

On real data, $G_T$ may be too large to store explicitly.
Ropebwt3 instead explicitly stores $G_P$ only and traverses it in the topological order (Algorithm~\ref{algo:bwa-sw}).
At a node $u\in V(G_P)$, we use a hash table to keep $\{v\in V(G_T):H_{uv}>0\}$.
This algorithm is exact in that it guarantees to find the best alignment.
In practice, however, a large number of $v\in V(G_T)$ may be aligned to $u$ with $H_{uv}>0$.
It is slow and memory demanding to keep track of all cells $(u,v)$ with positive scores when $P$ is long.
Similar to our earlier work, we only store top $W$ cells (25 by default) at each $u$.
This heuristic is akin to dynamic banding for linear sequences~\citep{Suzuki:2018aa}.

\begin{algorithm}[tb]
	\caption{The revised BWA-SW algorithm}\label{algo:bwa-sw}
	\begin{algorithmic}[1]
		\Procedure{BwaSW}{$G_P,G_T$}
			\For{$u\in V(G_P)$ in topological order}
				\For{$u'\in{\rm pre}(u)$}\Comment{predecessors of $u$}
					\For{$v'\in V(G_T)$ \emph{s.t.} $H_{u'v'}>0$}\Comment{match}
						\For{$v\in{\rm child}(v')$}\Comment{children of $v'$}
							\State $H_{uv}\gets \max\{H_{uv},H_{u'v'}+s(u',u;v',v)\}$
						\EndFor
					\EndFor
					\For{$v\in V(G_T)$ \emph{s.t.} $H_{u'v}>0$}\Comment{insertion}
						\State $E_{uv}\gets\max\{E_{uv},\max\{H_{u'v}-q,E_{u'v}\}-e\}$
						\State $H_{uv}\gets\max\{H_{uv},E_{uv}\}$
					\EndFor
				\EndFor
				\For{$v'\in V(G_T)$ \emph{s.t.} $H_{uv'}>0$}\Comment{deletion}
					\For{$v\in{\rm child}(v')$}\Comment{children of $v'$}
						\State $F_{uv}\gets\max\{F_{uv},\max\{H_{uv'}-q,F_{uv'}\}-e\}$
						\State $H_{uv}\gets\max\{H_{uv},F_{uv}\}$
					\EndFor
				\EndFor
			\EndFor
		\EndProcedure
	\end{algorithmic}
\end{algorithm}

\subsection{Estimating local haplotype diversity}

BWA-SW with the banding heuristic may miss the best matching haplotype especially given an index consisting of similar haplotypes.
When the suffix on the best full-length alignment has a lot more mismatches than the suffix on suboptimal alignments,
the best alignment may have moved out of the band early in the iteration and thus get missed.
Nevertheless, a long read sequenced from a new sample may be the recombinant of two genomes in the index.
We often do not seek the best alignment of the long read to a single haplotype.
We are instead more interested in the collection of haplotypes a query sequence can be aligned to even if they do not lead to the best alignment.
This now becomes possible as BWA-SW explores suboptimal alignments to multiple haplotypes.

More exactly, we perform semi-global sequence-to-DAWG alignment (i.e. requiring the query sequence to be aligned from end to end)
by applying BWA-SW to the graph representing the linear query sequence $P$, from the end to the start.
We can find the set of matching haplotypes $\mathcal{M}\triangleq\{v\in V(G_T):H_{u_0v}>0\}$ where $u_0$ represents the start of $P$.
$\mathcal{M}$ may include suboptimal haplotypes caused by small variants as well as the optimal one.

Importantly, $P$ may be aligned to similar positions on $T$ with slightly different gap placements.
For example, given $T={\tt CAAGCAG}$ and $P={\tt AGCG}$,
the algorithm above may find the following two alignments around the same position but in different SA intervals:
\begin{verbatim}
   T: CAAGCAG        CAAGCAG
        ||||    or    | |||
   P:   AGCA          A-GCA
\end{verbatim}
When counting hits, we want to ignore the second suboptimal alignment.
In theory, we can identify this case by comparing the position of each alignment.
This procedure is slow as the locate operation is costly.
We instead leverage the bi-directionality to identify redundancy heuristically.
Suppose $P$ can be aligned to SA bi-intervals $(k_1,k'_1,s_1)$ and $(k_2,k'_2,s_2)$ and the alignment score to the first interval is higher.
We filter out the second interval if $[k_2,k_2+s_2)\subset[k_1,k_1+s_1)$ or $[k'_2,k'_2+s_2)\subset[k'_1,k'_1+s_1)$.
This strategy does not avoid all overcounting but it works well on multiple real examples we have closely inspected.

In practice, we may apply this algorithm to the flanking sequence of a variant
or to sliding k-mers of a long query sequence to enumerate possible local haplotypes and estimate their frequencies in the index.
It is a new query type that is biologically meaningful.

\section{Results}

\begin{table}[!tb]
\caption{Datasets\label{tab:data}}
\begin{tabular*}{\columnwidth}{@{\extracolsep\fill}lrrr@{\extracolsep\fill}}
\toprule
Name               & \#bases$^1$ & \#sequences & avg run length$^2$ \\
\midrule
human100$^3$       &  301.6 Gb &  38.6 k & 141.6 \\
human320$^4$       &  963.0 Gb &  27.1 k & 395.6 \\
CommonBacteria$^5$ & 7326.6 Gb & 278.4 M & 828.6 \\
\botrule
\end{tabular*}
\begin{tablenotes}\setlength\itemsep{0.0em}
\item[$^{1}$] number of bases in the input sequences on one strand
\item[$^{2}$] average run length in BWT constructed from both strands
\item[$^{3}$] 100 long-read human assemblies; N50 string length: 44.4 Mb
\item[$^{4}$] 320 long-read human assemblies; N50 string length: 135.3 Mb
\item[$^{5}$] AllTheBacteria~\citep{Hunt2024.03.08.584059} excluding ``dustbin'' and ``unknown''
\end{tablenotes}
\end{table}

\begin{table}[b]
\caption{Indexing performance\label{tab:index}}
\begin{tabular*}{\columnwidth}{@{\extracolsep\fill}llrrr@{\extracolsep\fill}}
\toprule
Dataset        & Algorithm     & Elapsed$^1$ & CPU$^2$  &      RAM \\
\midrule
human100       & grlBWT        &  8.3 h       &  $\times3.6$ &  84.8 GB \\
		       & pfp-thres$^3$ &$<$51.7 h     &  $\times1.0$ & 788.1 GB \\
               & ropebwt3      & 20.5 h       & $\times22.9$ &  83.0 GB \\
			   & metagraph$^4$ & 16.9 h       & $\times18.6$ & 251.0 GB \\
			   & fulgor$^5$    & 1.2 h        & $\times27.2$ & 165.2 GB \\
human320       & grlBWT$^6$    & 23.3 h       &  $\times4.2$ & 270.4 GB \\
               & ropebwt3      & 81.2 h       & $\times16.5$ &  99.2 GB \\
               & ropebwt3$^7$  & 64.9 h       & $\times23.7$ & 170.5 GB \\
CommonBacteria & ropebwt3      & 25.6 d       & $\times32.4$ &  67.3 GB \\
\botrule
\end{tabular*}
\begin{tablenotes}\setlength\itemsep{0.0em}
\item Up to 64 threads specified if multi-threading is supported.
\item[$^{1}$] excluding time for format conversion; ``h'' for hours; ``d'' for days
\item[$^{2}$] number of CPU threads used on average
\item[$^{3}$] pfp-thresholds was run on a slower machine with more RAM
\item[$^{4}$] k-mer coordinates in the ``row\_diff\_brwt\_coord'' encoding
\item[$^{5}$] without ``-{}-meta -{}-diff'' as the basic index is smaller; lossy index
\item[$^{6}$] using two bytes per run with option `-b 2', which is faster
\item[$^{7}$] BWT merge and partial BWT construction with libsais are run together
\end{tablenotes}
\end{table}

\subsection{Performance on index construction}

We evaluated the performance of BWT construction on 100 haplotype-resolved human assemblies collected in~\citet{Li:2024ab}.
As we included both strands (Section~\ref{sec:ds-bwt}), each BWT construction algorithm took about 600 billion bases as input (Table~\ref{tab:data}).
grlBWT (commit 5b6d26a; \citealt*{DBLP:journals/iandc/DiazDominguezN23}) is the fastest algorithm (Table~\ref{tab:index})
at the cost of $\sim$2 terabytes of working disk space including decompressed sequences.
Ropebwt3 took 21 hours from input sequences in gzip'd FASTA to the final index, of which 7.7 hours was spent on libsais.
It does not use working disk space and can append new sequences to an existing BWT.
However, hardcoded for the DNA alphabet, ropebwt3 does not work with more general alphabets.
pfp-thresholds~\citep{Rossi:2022aa} used more memory than the input sequences.
It may be impractical with increased sample size.

Both MONI (v0.2.1; \citealt*{Rossi:2022aa}) and Movi (v1.1.0; \citealt*{Zakeri:2024aa}) use pfp-thresholds for building BWT.
They generate additional data structures on top of BWT.
Time spent on these additional steps were not counted.
The tested version of Movi used more than one terabyte of memory to construct the final index.
The Movi developer kindly provided the Movi index for the evaluation of query performance in the next section.

We also indexed the same dataset with k-mer based tools.
In the lossless mode, metagraph (v0.3.6; \citealt*{Karasikov2020.10.01.322164}) indexed the 100 human genomes in 17 hours (Table~\ref{tab:index}).
Fulgor (v3.0.0; \citealt*{Fan:2024aa}) is by far the fastest.
However, lossy in nature, it is not directly comparable to the rest of the tools in the table.

To test scalability, we indexed a larger dataset consisting of 320 human genomes recently released by the Human Pangenome Reference Consortium~\citep{Liao:2023aa}.
Because these assemblies are more contiguous than human100 (Table~\ref{tab:data}), $m_2$ in Algorithm~\ref{algo:merge} is smaller, which reduces the multi-threading efficiency of ropebwt3.
We can alleviate this by executing BWT merge and partial BWT construction at the same time at the cost of higher memory footprint.
On human320, grlBWT is 2--4 times as fast but uses more memory (Table~\ref{tab:index}).
On the largest CommonBacteria dataset (Table~\ref{tab:data}),
ropebwt3 took less than a month to construct the double-strand BWT with 14.66 trillion symbols.
The final index in the fermi format is 27.6 GB in size.
The ``dustbin'' and ``unknown'' categories in AllTheBacteria consist of many unique sequences which are not compressed well.
Indexing the entire AllTheBacteria with ropebwt3 will probably take 100--200 GB memory at the peak.

\begin{table}[!tb]
\caption{Query performance\label{tab:query}}
\begin{tabular*}{\columnwidth}{@{\extracolsep\fill}lllrr@{\extracolsep\fill}}
\toprule
Data   & Algorithm     &Type$^4$&Speed$^5$ (kb/s)&RAM (GB) \\
\midrule
SR$+^1$& ropebwt3$^6$  & MEM31  & 1758.5       &  10.6 \\
       &               & MEM51  & 1907.5       &  10.6 \\
	   &               & SW10   & 84.1         &  15.2 \\
%	   &               & suffix & 2216.4       &  10.3 \\
	   & MONI$^7$      & MEM$-$ & 453.2        &  68.4 \\
	   &               & extend & 348.2        &  68.4 \\
       & Movi          & PML    & 5894.0       &  47.6 \\
%	   &               & suffix & 6792.0       &  79.4 \\
	   & metagraph     & PA$+$  & $<$0.1       &  65.3 \\
	   & fulgor        & PA     & 2717.5       &   5.1 \\
LR$+^2$& ropebwt3      & MEM31  & 1695.9       &  10.5 \\
       &               & MEM51  & 1793.9       &  10.5 \\
       &               & SW25   & 82.7         &  15.6 \\
	   & MONI          & MEM$-$ & 413.6        &  68.4 \\
	   & Movi          & PML    & 16204.9      &  47.6 \\
	   & metagraph     & PA$+$  & $<$0.1       &  65.3 \\
	   & fulgor        & PA     & 2491.6       &   5.1 \\
LR$-^3$& ropebwt3      & MEM31  & 1365.0       &  10.4 \\
       &               & MEM51  & 3051.6       &  10.4 \\
	   &               & SW25   & 58.2         &  17.9 \\
	   & MONI          & MEM$-$ & 186.8        &  68.4 \\
	   & Movi          & PML    & 8490.9       &  47.6 \\
	   & metagraph     & PA$+$  & 1119.3       &  65.3 \\
	   & fulgor        & PA     & 4240.8       &   5.1 \\
\botrule
\end{tabular*}
\begin{tablenotes}\setlength\itemsep{0.0em}
\item[$^1$] first 1 million 125bp human short reads from SRR3099549
\item[$^2$] first 10,000 human PacBio HiFi reads from SRR26545347
\item[$^3$] first 10,000 metagenomic PacBio HiFi reads from DRR290133
\item[$^4$] MEM$x$: super-maximal exact matches (SMEMs) of $x$ bp or longer with counts;
%suffix: longest matching suffix with counts;
MEM$-$: SMEM without counts;
extend: Smith-Waterman extension from the longest SMEM;
PML: pseudo-matching length; PA: pseudo-alignment; PA$+$: pseudo-alignment with contig names;
SW$y$: BWA-SW with bandwidth $y$
\item[$^5$] kilobases processed per CPU second. Index loading time excluded
\item[$^6$] index in the binary fermi format
\item[$^7$] The MONI index includes both strands.
We modified MONI such that extension is performed on the forward query strand only
\end{tablenotes}
\end{table}

\subsection{Query performance}

We queried 100--200 Mb human short reads (SR$+$), human long reads (LR$+$) and non-human long reads (LR$-$) against the human pangenome indices constructed above (Table~\ref{tab:query}).
It is important to note that no two tools support exactly the same type of query,
but the comparison can still give a hint about the relative performance.

Among the three BWT-based tools, ropebwt3 is slower than Movi but faster than MONI on finding exact matches.
Movi finds pseudo-matching length (PML) which is not intended to be the longest exact match.
PML corresponds to the longest MEM for only 195 out of one million short reads, and none for long reads.
The longest PML of each read is on average 27\% shorter than the longest exact match for LR$+$ and 22\% shorter for SR$+$.
Nonetheless, we believe it is possible to implement SMEM finding based on the Movi data structure with minor performance overhead.

MONI and ropebwt3 can also find inexact matches.
Implementing r-index, MONI can relatively cheaply locate one SMEM.
It leverages this property to extract the genomic sequence around the longest SMEM and performs Smith-Waterman extension.
MONI extension is faster than BWA-SW on SR$+$ because it does not need to inspect suboptimal hits.
This feature apparently focuses on short reads only.
On LR$+$, MONI extension fails to extend to the ends of reads and generate incorrect output for the majority of reads.

As to k-mer indices, fulgor outputs the labels of genomes that each read have enough 31-mer matches to.
In its current form, such output is not useful for pangenome analysis as we know most of reads can be mapped to all genomes.
Metagraph can additionally output the contig name of each match.
However, when most k-mers are present in the index, metagraph is impractically slow.

\subsection{Identifying novel sequences}

As a biological application, we used the pangenome index to identify novel sequences in reads that are absent from other human genomes.
For this, we downloaded the PacBio HiFi reads for tumor sample COLO829 (\url{https://downloads.pacbcloud.com/public/revio/2023Q2/COLO829/COLO829/}),
mapped them to human100 (Table~\ref{tab:data}), which includes T2T-CHM13~\citep{Nurk:2022up},
and extracted $\ge$1kb regions on reads that are not covered by SMEMs of 51bp or longer.
We found 95kb sequences in 43 reads.
These sequences could not be assembled.
NCBI BLAST suggested multiple weak hits to cow genomes.
We could not identify the source of these sequences but there were few of them anyway.

When we mapped the COLO829 reads to T2T-CHM13 only and applied the same procedure,
we found 55.9Mb of ``novel'' sequences in 25,365 reads.
The much larger number is caused by regions with dense SNPs that prevented long exact matches.
Counterintuitively, mapping these reads to T2T-CHM13 with minimap2~\citep{Li:2018ab} resulted in more ``novel'' sequences at 78.6Mb,
probably because minimap2 ignores seeds occurring thousands of times in T2T-CHM13 and may miss more alignments.
Capable of finding SMEMs at the pangenome scale, ropebwt3 is more effective for identifying known sequences.
It is also 16\% faster and uses less memory than full minimap2 alignment against a single genome.

\subsection{Haplotype diversity around C4 genes}

\begin{figure}[tb]
\includegraphics[width=.49\textwidth]{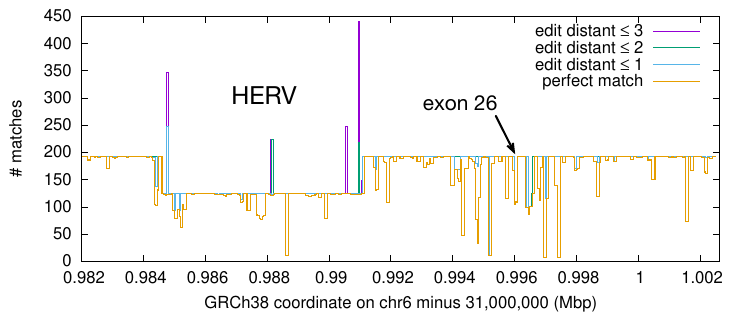}
\caption{Haplotype diversity around the \emph{C4A} gene.
101-mers with 50bp overlaps are extracted from the \emph{C4A} genomic sequence on GRCh38
and aligned to the human pangenome.
The Y axis shows the counts of 101-mer matches under different edit-distance thresholds.}\label{fig:c4}
\end{figure}

The reference human genome GRCh38 has two paralogous C4 genes, \emph{C4A} and \emph{C4B}, and they may have copy-number changes~\citep{Sekar:2016aa}.
In both cases, the exon 26 harbors the functional domain.
We extracted the exon 26 from both genes from GRCh38.
They differ at six mismatches over 157bp.
We aligned both 157bp segments, one from each gene, to the human pangenome.
105 haplotypes have the same \emph{C4A} exon 26 sequence, one haplotype has a mismatch at offset 128 and four haplotypes have a mismatch at 54.
In case of \emph{C4B}, 60 haplotypes have the same reference \emph{C4B} sequence and 23 also have a mismatch at offset 54.
This means among the 100 human haplotypes, there are 110 \emph{C4A} gene copies and 83 \emph{C4B} copies.
If we increase the bandwidth from the default 25 to 200, BWA-SW will be able to align \emph{C4A} exon 26 to \emph{C4B} genes
and output all five hits for each sequence.

Fig.~\ref{fig:c4} shows the local haplotype diversity across the entire \emph{C4A} gene spanning $\sim$20.6kb on GRCh38.
We can see most regions have 193 copies, except a $\sim$6.4kb HERV insertion that separates long and short forms~\citep{Sekar:2016aa}.
The dip around exon 26 is caused by the \emph{C4A}--\emph{C4B} difference.
We could only see these alternative haplotypes with BWA-SW which reports suboptimal hits.

\section{Discussions}

Ropebwt3 is a fast tool for BWT construction and sequence search for redundant DNA sequences.
Generating BWT purely in memory and supporting incremental build, ropebwt3 is convenient and practical for BWT construction at large scale.
It is the only algorithm that can construct the BWT of 320 human genomes from a 2.5 GB AGC archive~\citep{Deorowicz:2023ab} without staging all decompressed data in memory or on disk.
It provides the fastest algorithm so far for finding supermaximal exact matches and can report inexact hits as well.
Ropebwt3 demonstrates that BWT-based data structures are scalable to terabases of pangenome data.

%Ropebwt3 is likely to take $\sim$3 weeks to index 1,000 human genomes.
%When there are more genomes, it will be faster to construct the BWT of individual genome independently
%and then merge them one by one.
%This strategy will use working disk space but achieve better parallelization.
%\citet{DBLP:conf/dcc/Siren16} introduced disk-based algorithms for merging large BWTs, which may further improve the parallelization of BWT merge.

Ropebwt3 implements an FM-index to locate SMEMs or local hits.
Although the standard r-index is faster than an FM-index of the same size,
it imposes a fixed sampling rate: two suffix array values per run.
The BWT of CommonBacteria has 14.6 trillion bases and 17.6 billion runs.
An r-index is likely to take more than 200GB,
while an FM-index sampled at one position per 8,192bp takes 17.5GB in ropebwt3.
Subsampled r-index (sr-index; \citealt*{DBLP:journals/corr/abs-2409-14654}) is probably the better solution, which we will explore in future.

This article has not evaluated several recent BWT-based tools including
r-index-f~\citep{DBLP:conf/wea/BrownG022}, block\_RLBWT~\citep{DBLP:conf/wea/Diaz-DominguezD23}, Move-r~\citep{DBLP:conf/wea/Bertram0N24} and b-move~\citep{DBLP:conf/wabi/DepuydtRVVGF24}.
Although they have not been tested on large datasets like human pangenome and they do not report SMEMs,
some of their data structures are probably more efficient than the ones we use.
We may integrate these methods into ropebwt3 in future as well.

We often use pangenome graphs to analyze multiple similar genomes~\citep{Liao:2023aa}.
These graphs are built from the multiple sequence alignment through complex procedures involving many parameters~\citep{Li:2020aa,Hickey:2023aa,Garrison:2024aa}.
It is challenging to understand if the graph topology is biologically meaningful especially given that we do not know the correct alignment between two genomes, let alone multiple ones.
Complement to graph-based data structures, BWT-based algorithms are often exact with no heuristics or parameters but they tend to support limited query types.
For example, we cannot project the alignment to a designated reference genome.
What additional query types we can achieve will be of great interest to the comprehensive pangenome analysis in future.

\section*{Acknowledgments}

We are grateful to Travis Gagie for pointing us to his long MEM finding algorithm,
to Ilya Grebnov for adding the support of 16-bit alphabet which helps to accelerate ropebwt3,
to Mohsen Zakeri for providing the Movi index,
to Massimiliano Rossi for explaining the MONI algorithm,
and to Giulio Pibiri and Rob Patro for trouble-shooting compilation issues with fulgor.

\section*{Author contributions}

H.L. conceived the project, implemented the algorithms, analyzed the data and drafted the manuscript.

\section*{Conflict of interest}

None declared.

\section*{Funding}

This work is supported by National Institute of Health grant R01HG010040 and U01HG010961 (to H.L.).

\section*{Data availability}

The ropebwt3 source code is available at \url{https://github.com/lh3/ropebwt3}.
Prebuilt ropebwt3 indices can be obtained from \url{https://doi.org/10.5281/zenodo.11533210}
and \url{https://doi.org/10.5281/zenodo.13955431}.

\bibliographystyle{apalike}
{\sffamily\small
\bibliography{ropebwt3}}

\end{document}